\documentclass{ws-procs9x6square}
\usepackage{amsmath,amssymb}

\newif\ifuseprd
\useprdfalse
\newif\ifeprint
\eprinttrue
\newif\ifdatelast
\datelasttrue
\newif\iftoomuchdetail
\toomuchdetailfalse

\newcommand{\eqn}[1]{(\ref{#1})}
\newcommand{\be}{\begin{equation}}
\newcommand{\ee}{\end{equation}}
\newcommand{\ben}{\begin{displaymath}}
\newcommand{\een}{\end{displaymath}}
\newcommand{\bea}{\begin{eqnarray}}
\newcommand{\eea}{\end{eqnarray}}
\newcommand{\bean}{\begin{eqnarray*}}
\newcommand{\eean}{\end{eqnarray*}}

\newcommand{\ba}{\begin{array}}
\newcommand{\ea}{\end{array}}
\newcommand{\ei}{\end{itemize}}

\def\l {\lambda}
\def\a {\alpha}

\def\G {\Gamma}

\def\s {\sigma}
\def\e {\epsilon}

\def\otaula{\begin{tabular}}
\def\ctaula{\end{tabular}}

\renewcommand{\t}{\theta}
\def\CR{\mathbb{R}}
\def\CM{\mathcal{M}}

\def\s{\sigma}
\def\8M{$\CM_8$}

\def\be{\begin{equation}}
\def\ee{\end{equation}}
\def\G{\Gamma}

\def\ei{e^{\underline{i}}}

\def\e1{e^{\underline{1}}}
\def\1u{\underline{1}}
\def\2u{\underline{2}}

\def\0u{\underline{0}}
\def\e{\epsilon}
\def\target{$\CR^{1,1}\times \mathcal{M}_8$ }
\def\target2{$\CR^{1,1}\times \mathcal{M}_8$,}
\def\9G{\G_{\underline{9}}}

\def\a{\alpha}

\newcommand{\caln}{\mbox{${N}$}}

\newcommand{\ads}[1]{\mbox{${AdS}_{#1}$}}
\newcommand{\adss}[2]{\mbox{$AdS_{#1}\times {S}^{#2}$}}

\newcommand{\fc}{\frac}

\newcommand{\sac}{\, , \qquad}

\def\math@strike#1#2{%
  \setbox\z@\hbox{$\m@th#1{#2}$}\fin@strike}
\def\make@strike#1{%
  \setbox\z@\hbox{\color@begingroup#1\color@endgroup}\fin@strike}
\def\fin@strike{%
  \@tempdima\dp\z@
  \@tempdimb\ht\z@
  \lower\@tempdima\hbox{\strike@start}%
  \box\z@
  \raise\@tempdimb\hbox{\strike@end}}
\def\strike@start{\special{ps: %
    currentpoint /starty exch def /startx exch def}}
\def\strike@end{
}
\newcommand\fs{\protect\@strike}

\ifeprint\def\draftnote{}\fi

\begin{document}


\title{
\ifeprint
\vspace{-3\baselineskip}
\begingroup
\footnotesize\normalfont\raggedleft
\lowercase{\sf hep-th/0401058 \\} 
\vspace{\baselineskip}
\endgroup
\fi
More on supersymmetric tensionless 
rotating strings in $A\lowercase{d}S_5\times S^5$%
\ifeprint
\footnote{
\uppercase{T}o appear in the \uppercase{P}roceedings of
the 3rd 
\uppercase{I}nternational \uppercase{S}ymposium on \uppercase{Q}uantum
\uppercase{T}heory and \uppercase{S}ymmetries (\uppercase{QTS}3), 
\uppercase{C}incinnati, 10~--~14 \uppercase{S}ept 2003 ---
\copyright\
\uppercase{W}orld \uppercase{S}cientific.}
\fi
}


\author{DAVID MATEOS}
\address{Perimeter Institute for Theoretical Physics \\
Waterloo, Ontario N2J 2W9, Canada \\
E-mail: dmateos@perimeterinstitute.ca}

\author{TONI MATEOS}
\address{Departament ECM, Facultat de F\'\i sica \\
Universitat de Barcelona, Institut de F\'\i sica d'Altes Energies and \\
CER for Astrophysics, Particle Physics and Cosmology \\
Diagonal 647, E-08028 Barcelona, Spain \\
E-mail: tonim@ecm.ub.es}

\author{PAUL K. TOWNSEND\footnote{\uppercase{O}n leave from 
\uppercase{DAMTP}, \uppercase{U}niversity of \uppercase{C}ambridge, 
\uppercase{UK}.}}
\address{Instituci\'o Catalana de Recerca i Estudis Avan\c cats.\\
Departament ECM, Facultat de F\'\i sica \\
Universitat de Barcelona \\ 
Diagonal 647, E-08028 Barcelona, Spain \\
E-mail: p.k.townsend@damtp.cam.ac.uk}


\maketitle

\abstracts{
Rotating IIB strings in $AdS_5 \times S^5$ become ultra-relativistic, 
and hence effectively tensionless, in the limit of large angular momentum 
on $S^5$. We have shown previously that such tensionless strings may 
preserve supersymmetry. Here we extend this result to include a class 
of supersymmetric tensionless strings with {\it arbitrary} $SO(6)$ 
angular momentum. We close with some general comments on tensionless
strings in $AdS_5 \times S^5$.}


\section{Introduction}

A recent advance in our understanding of the AdS/CFT correspondence
has come about via consideration of classical, closed IIB strings in 
\adss{5}{5} that rotate within $S^5$ and thus carry 
$SO(6)$ momenta \cite{FT12,FT3}. In the limit of large angular momentum
$J$ (precisely, for $J \gg 1, \sqrt{\l}$, where $\sqrt{\l}$ is the 
string tension) the energy of the string is given  
by the classical result, and this admits an expansion in positive integer
powers of $\l /J^2 $. This yields a prediction for
the dimension of the dual operator, with corresponding large $SO(6)$ 
R-charge, in {\it perturbative} $\caln=4$ super Yang-Mills (SYM) 
theory. In the planar limit, this dimension can be computed by spin 
chain techniques \cite{BMSZ} (or by means of an equivalent sigma model
\cite{Martin}) and a precise agreement with the string 
theory predictions has been found; remarkably, both routes lead to the 
same integrable model \cite{AFRT,BFST,AS,EMZ}. 

The cases that were first studied, and which are still best understood, 
are those for which the $SO(6)$ angular momentum two-form has only 
two {\it independent} skew-eigenvalues. It was argued in \cite{FT12,FT3}  
that the agreement between the string and SYM calculations was, 
in contrast to previous quantitative tests, evidence for the AdS/CFT 
conjecture in a `non-supersymmetric sector'.
However, as we pointed out in a previous paper 
\cite{MMT}, the rotating strings under consideration are
effectively tensionless for large angular momentum because almost all
of the energy is due to the rotation, and the states of tensionless
strings in the same $SO(6)$ representation preserve either 1/4 or 1/8 
supersymmetry. Thus, the limit in which the AdS/CFT prediction can be 
checked by semiclassical, perturbative techniques on both sides of 
the correspondence is a limit in which supersymmetry is partially 
restored. 

More recently, predictions of the AdS/CFT correspondence have been 
verified for a larger class of rotating strings for which the $SO(6)$ 
angular momentum has three independent skew-eigenvalues
\cite{ART}.  One finds that the energies, like the corresponding
anomalous dimensions on the $\caln=4$ SYM  side of the correspondence,  
are again determined by the solution to an integrable model, but of a 
more general type. Here we shall show, extending our previous result,  
that  these rotating strings  are again supersymmetric in the 
ultra-relativistic, and hence tensionless, limit. 
In fact, the tensionless strings  are supersymmetric for {\it
arbitrary} $SO(6)$ angular momentum, generically preserving 1/8 
supersymmetry. 

This paper is written with the intention that it be read in
conjunction with \cite{MMT}. We will begin by presenting the tensionless 
version of the general `three-spin' rotating string solution of
\cite{ART}. We then present our main new result, which is a
demonstration that these tensionless strings generically preserve 
1/8 supersymmetry. We close with some further comments on tensionless
strings in \adss{5}{5}.

\section{Null rotating strings on $S^5$}

Since the strings of interest here lie
at the origin of \ads{5}, the relevant part of the \adss{5}{5} metric
is
\be
\label{relevant}
ds^2= -dt^2 + d\t^2+\sin^2 \t d\phi^2 + 
\sum_{i=1}^3 a_i^2 (d\a_i)^2 \,,
\ee
where $a_1 = \cos \t$,  $a_2=\sin\t \cos\phi$, $a_3=\sin\t \sin \phi$,
and $\a_i$ are polar angles in three orthogonal planes.
In order to avoid coordinate singularities, we will assume here that
$a_1 a_2 a_3$ is non-zero, which corresponds to the assumption that 
the angular momentum two-form has maximal rank.

Let $(\tau,\s)$ be the worldsheet coordinates. In the gauge
$t=\tau$ the phase-space form of the Lagrangian density for a 
tensionless string in the above background is
\be
L = p_\t \dot \t + p_\phi \dot\phi + \sum_i p_i \dot \alpha_i - H
-s\left(p_\t \t'+p_\phi \phi'+ \sum_i p_i \alpha_i'\right) \,,
\ee
where $s$ is the Lagrange multiplier for the string reparametrization
constraint, and $H$ is the Hamiltonian density
\be
H= \sqrt{ p_\t^2 + \frac{p_\phi^2}{\sin\t^2} + 
\sum_i \frac{p_i^2}{a_i^2}} \,. 
\ee
Note that the $\alpha_i$ equation of motion is
$\dot p_i = \left(sp_i\right)'$, 
from which it follows that the angular momentum,
$J_i = \oint d\sigma \, p_i$, is a constant of motion. 
 
We seek solutions for which $\t$ and $\phi$ are constant and 
$p_\t=p_\phi=0$. This requires, in order to solve the 
corresponding equations of motion, that we set
\be
\label{one}
a_i^2 = |p_i| \Big/ \sum_j |p_j| \,,
\ee
from which we see that 
\be \label{ham}
H= \sum_i |p_i| \,.
\ee

Given \eqn{one}, the $p_i$ equations of motion reduce to
$\dot\alpha_i -s \alpha_i' = 1$, while the constraint imposed by 
$s$ is $\sum_i p_i \alpha'_i =0$.
We may choose a gauge for which  $s=0$, in which case the equations
above have the solution\footnote{This is not the unique solution, but
it is the one of relevance when one considers the tensionless string
as a limit of the tensionful string.}
\be
\alpha_i = t + m_i \sigma \sac p_i = J_i /2\pi \,,
\ee
for integers $m_i$  satisfying $\sum_i m_i J_i  = 0$. 
Because each $p_i$ has a definite sign, integration of \eqn{ham} yields
the total energy
\be
E = |J_1|+ |J_2|+ |J_3| \,.
\ee
An argument analogous to that of \cite{MMT} shows that this energy
saturates a  BPS bound implied by the $PSU(2,2|4)$ 
supersymmetry algebra of \adss{5}{5}. Because of this, 
the rotating strings are supersymmetric. Since we have assumed that
all $J_i$ are non-zero, the fraction of supersymmetry preserved is
1/8, as we will confirm in the following 
section. If only two $J$'s are non-zero, then 1/4 supersymmetry is
preserved. And if only one $J$ is non-zero, then 1/2 supersymmetry is
preserved.

\section{Supersymmetry}

The condition for a IIB superstring to be supersymmetric in 
the ultra-relativistic limit in which it becomes null is 
\be
p_M \, e^M_{\,\,\,\,\,\,A} \, \G^A \, \epsilon =0 \,,
\label{susy}
\ee
where $p_M$ is the ten-momentum, $e^M_{\,\,\,\,\,\,A}$ is the obvious 
orthonormal frame associated to the metric \eqn{relevant}, $\G^A$ are the 
corresponding tangent-space (constant) ten-dimensional Dirac 
matrices, and $\epsilon$ is a chiral, complex Killing spinor of 
the \adss{5}{5} background. This spinor can be written in terms of 
a constant spinor $\e_0$ as
\be
\e=e^{i \frac{t}{2} \tilde{\G}} e^{i \frac{\t}{2} \G_{\phi 123}} 
e^{\frac{\phi}{2}\G_{\t\phi}} e^{\fc{i}{2} \left( 
\a_1 \G_{\t\phi 23} + \a_2 \, i \G_{2 \t} + \a_3 \, i \G_{3\phi} 
\right)} \, \e_0 \,,
\label{epsilon}
\ee 
where $\tilde{\G}$ commutes with all other matrices in the problem
and $\G_i \equiv \G_{\a_i}$. It is clear from this expression that 
the matrices occurring inside the brackets in the last exponential 
generate, when acting on $\e_0$, rotations in each of the three 
orthogonal two-planes parametrised by $\a_i$. Note that they all 
square to unity (so they have eigenvalues $\pm 1$), they are 
mutually-commuting, and the product of any two of them yields the 
third one, up to a sign.

For the rotating string solutions of the previous section, the 
supersymmetry condition \eqn{susy} reduces to
\be
a_i \, \G_{ti} \, \epsilon = \epsilon \,,
\label{a}
\ee
where we have assumed, for definiteness, that all $J_i$ are positive. 
If the angular momentum two-form has maximum rank then all $a_i$ 
are non-zero, and equation \eqn{a} is equivalent to
\be
i\G_{2\t} \, \e_0 = \e_0 \sac 
i\G_{3\phi} \, \e_0 = \e_0 \sac 
\G_{t1} \, \e_0 = \e_0 \,.
\label{susybis}
\ee
Let us first show that \eqn{susybis} implies \eqn{a}. 
We see from the first two conditions in \eqn{susybis}
that $\e_0$ is an eigen-spinor of the last exponential in 
equation \eqn{epsilon}, so this exponential cancels on both sides 
of \eqn{a}. The first exponential also cancels because $\tilde{\G}$ 
commutes with all other matrices in equation \eqn{a}, so the latter 
may be rewritten as 
\be
a_i \, \G_{ti} \, \e_0 = e^{-\frac{\phi}{2}\G_{\t\phi}}
e^{i \t \G_{\phi 123}} e^{\frac{\phi}{2}\G_{\t\phi}} \, \e_0 =
\left( a_1 + a_2 \, i \G_{\phi 123} + 
a_3 \, i \G_{123\t} \right) \, \e_0 \,,
\ee
which is identically satisfied by virtue of \eqn{susybis}.

With some further algebra, it can also be shown that \eqn{a} implies
\eqn{susybis}, but we will not do this here. Instead, we note that, 
being mutually-commuting projections, the conditions \eqn{susybis} 
imply that the fraction of preserved supersymmetry is 1/8, 
as expected for a string carrying three independent angular 
momenta. We also recall that, as observed above, they imply that 
$\e_0$ is invariant (up to a phase) under rotations in the 
$\a_i$-planes associated to the non-zero components of the angular 
momentum two-form.

\section{Discussion}

There are now many string solutions in \adss{5}{5} whose
classical energies have been matched, in the large-energy limit, 
to anomalous dimensions of composite operators, with a large 
number of fields, in perturbative \mbox{$N=4$} SYM theory. 
{\it All} these strings become ultra-relativistic in this
limit, and hence effectively tensionless, in the sense that 
their kinetic energy, $K$, is much larger than the energy 
associated to their tension; their classical energy then admits an
expansion in positive powers of $\l / K^2$.\footnote{Some aspects
of these ultra-relativistic strings have been studied in 
\cite{Miha}.}

This property is, of course, necessary for a 
comparison to {\it perturbative} SYM to be possible, and is
a typical property of large-energy strings moving on $S^5$,
as opposed to $AdS_5$. The reason is that the size
of a generic string moving on $S^5$ is bounded above by 
the radius of the sphere, so in this case the kinetic energy 
always dominates in the large-energy limit. On the contrary, 
the size of a string moving in $AdS_5$, and hence the contribution 
of its tension to its energy, can grow without bound.
In other words, the large-energy limit for generic strings 
moving in $AdS_5$ is not an ultra-relativistic limit.
It is therefore not surprising that the energies of these 
strings cannot be precisely matched to dimensions of SYM
operators computed within perturbation theory.

Almost all known strings that become tensionless in the
large-energy limit also become supersymmetric, 
as we have shown in this article for the class of rotating 
string solutions of \cite{ART} with generic $SO(6)$ angular 
momenta. Of course, in all these cases the energy 
saturates a BPS bound in the large-energy limit, 
$E=K=|J_1| + |J_2| + |J_3|$, and so this limit is equivalent
to a large-angular-momentum limit. The only apparent exception
that we are aware of is a pulsating string solution, for
which the classical energy matches an anomalous dimension of a SYM
operator computed in perturbation theory despite the fact that this
energy is not close to saturating a BPS bound \cite{EMZ}.
All string solutions for which it has been checked that the
sigma-model quantum corrections to the classical energy vanish in the
large-energy limit \cite{FT12} are also those that happen to be
supersymmetric in this limit \cite{MMT}.
The match for the pulsating string is therefore mysterious,
since it relies on the neglect of these corrections in the
absence of supersymmetry. 
This might suggest that the key property 
behind the success of these tests of the AdS/CFT conjecture is the 
ultra-relativistic nature of the corresponding strings,
rather than the nearly-supersymmetric property emphasised in our
previous paper.

\section*{Acknowledgments}

We are grateful to Jorge Russo for useful discussions.
DM is partially supported by funds from NSERC of Canada.


\end{document}